\documentclass[prd,twocolumn,superscriptaddress,showpacs,nofootinbib,preprintnumbers]{revtex4}

\usepackage{amsmath}
\usepackage{amsfonts}
\usepackage{graphicx}
\usepackage{dcolumn}

\def\be{\begin{equation}}
\def\ee{\end{equation}}
\def\ba{\begin{eqnarray}}
\def\ea{\end{eqnarray}}
\def\bs{\begin{subequations}}
\def\es{\end{subequations}}
\def\t{\tilde}

\newcommand{\rd}{{\rm d}}
\newcommand{\rde}{{\rm DE}}

\usepackage{color}

\begin{document}

\title{Matter density perturbations and effective gravitational constant \\
in modified gravity models of dark energy}

\author{Shinji Tsujikawa}
\email{shinji@nat.gunma-ct.ac.jp}
\affiliation{Department of Physics, Gunma National College of
Technology, Gunma 371-8530, Japan}
\date{\today}

\begin{abstract}
 
We derive the equation of matter density perturbations on sub-horizon scales
for a general Lagrangian density $f(R, \phi, X)$ that is a function of 
a Ricci scalar $R$, a scalar field $\phi$ and a kinetic term 
$X=-(\nabla \phi)^2/2$.
This is useful to constrain modified gravity dark energy models
from observations of large-scale structure and weak lensing.	
We obtain the solutions for the matter perturbation $\delta_m$ as 
well as the gravitational potential $\Phi$ for some analytically 
solvable models. In a $f(R)$ dark energy model with the Lagrangian 
density $f(R)=\alpha R^{1+m}-\Lambda$, the growth rates of 
perturbations exhibit notable differences from those in 
the standard Einstein gravity unless $m$ is very close to 0.
In scalar-tensor models with the Lagrangian 
density $f=F(\phi)R+2p(\phi,X)$ we relate the models with 
coupled dark energy scenarios in the Einstein frame 
and reproduce the equations of perturbations known in the current 
literature by making a conformal transformation.
We also estimate the evolution of perturbations in both Jordan and 
Einstein frames when the energy fraction of dark energy is constant 
during the matter-dominated epoch.
	
\end{abstract}


\maketitle

\section{Introduction}

Recent observations have determined basic cosmological parameters 
in high-precisions, but at the same time they posed a serious problem 
about the origin of dark energy (DE).
The analysis of Super-Nova Ia (SNIa) \cite{SN} is based upon the background 
expansion history of the universe around the redshift $z<{\cal O}(1)$.
The constraint obtained from SNIa so far has a degeneracy in 
the equation of state (EOS) of DE \cite{SN2}.   
To many people's frustration, the $\Lambda$CDM model with 
an EOS $w_{\rm DE}=-1$ has been continuously favored 
from observations. This degeneracy has been present
even adding other constraints coming from Cosmic Microwave 
Background (CMB) \cite{CMB} and Baryon Acoustic 
Oscillations (BAO) \cite{BAO}.

The models of dark energy can be broadly classified
into two classes \cite{review,CST}. The first corresponds 
to introducing a specific matter that leads to an accelerated expansion.
Most of scalar field models such as quintessence \cite{quin} and 
k-essence \cite{kes} belong to this class. 
The second class corresponds to so-called modified 
gravity models such as $f(R)$ gravity \cite{fR}, 
scalar-tensor theories \cite{st} and braneworld 
models \cite{brane}.
In order to break the degeneracy of observational constraints 
on $w_{\rm DE}$ and to discriminate between a host of 
DE models, it is important to find additional information 
other than the background expansion history of the Universe.
In this paper we will show that modified gravity models 
can be distinguished from others
by considering the evolution of 
matter perturbations $\delta_m$ and gravitational 
potentials $\Phi$.

In Einstein gravity it is well known that linear matter perturbations 
on sub-horizon scales satisfy the following equation 
\begin{eqnarray}
\ddot{\delta}_m+2H\dot{\delta}_m
-4\pi G \rho_m \delta_m=0\,,
\end{eqnarray}
where $H$ is a Hubble parameter, $G$ is a Newton's gravitational 
constant, $\rho_m$ is an energy density of the non-relativistic
matter, and a dot represents a derivative with respect to 
cosmic time $t$. During the matter-dominated epoch this has 
a growing-mode solution $\delta_m \propto a \propto t^{2/3}$,
which leads to the formation of large-scale structure.
In modified gravity models the growth rates of perturbations 
are different because of the modification of the gravitational 
constant as well as the change of the background evolution.
In the context of $f(R)$ gravity, in particular, there have been 
a number of recent works about the evolution of 
density perturbations during the matter-dominated and 
dark energy dominated epochs \cite{fRper}.

We will derive the equation of matter perturbations in Sec.~III 
for a very general Lagrangian density $f(R, \phi, X)$, where 
$R$ is a Ricci scalar and $\phi$ is a scalar field with 
a kinetic term $X=-(\nabla \phi)^2/2$.
Together with a sub-horizon approximation
we assume that $F\equiv \partial f/\partial R$
depends on $\phi$ and $R$ but not on $X$.
In fact this Lagrangian covers most of modified gravity 
DE scenarios such as $f(R)$ gravity models 
and scalar-tensor theories. The effect of modified gravity 
appears in an effective gravitational constant 
$G_{\rm eff}$ whose explicit form is given in 
Eq.~(\ref{Geff}). We derive a parameter $\eta$
introduced in Ref.~\cite{AKS} to quantify the 
strength of an anisotropic stress and also 
evaluate a parameter $\Sigma=q(1+\eta/2)$, where $q$ is 
a quantity that characterizes the deviation from the 
gravitational constant measured in solar system 
experiments today.
 
The results in this paper can be important for future surveys
of weak lensing \cite{lensing} as well as for the 
observations of large-scale structure (LSS) \cite{LSS}. 
In Ref.~\cite{Shirata} the deviation from Einstein gravity 
was constrained from the galaxy clustering by taking into account 
an additional Yukawa correction to the gravitational constant.
It will be possible to carry out similar observational constraints
on our $f(R,\phi, X)$ DE models from the LSS data by solving the 
equation of matter perturbations. 
In Ref.~\cite{AKS} the authors
proposed a DE parametrization using the variables 
$(\Sigma, \eta)$ together with a linear perturbation
growth factor $\gamma$ introduced in 
Refs.~\cite{Lahav,WS,Linder}.
If the deviation from the Einstein gravity case
$(\Sigma, \eta)=(1,0)$ is detected from future 
survey of weak lensing, this allows us to distinguish
modified gravity models from the models in Einstein gravity.

In Sec.\,IV we will find solutions for $\delta_m$ and $\Phi$ during the 
matter-dominated epoch for some analytically solvable models.
In particular we show that $f(R)$ dark energy models have a peculiar 
scale-dependence of perturbations unlike the case of Einstein gravity.
The effect of modified gravity on perturbations is important
provided that a dimensionless variable $m=Rf_{,RR}/f_{,R}$, 
which characterizes the deviation from the $\Lambda$CDM model, 
is not very close to 0.

The scalar-tensor models with the Lagrangian density
$f=F(\phi)R+2p(\phi, X)$ correspond to coupled dark energy 
models in the Einstein frame with a coupling $Q(\phi)=-F_{,\phi}/2F$.
In Sec.\,V we derive the equation of 
matter perturbations in the Einstein frame under a conformal transformation
and show that this in fact coincides with the equation 
in the models of dark energy coupled to the matter \cite{Ameper}. 
We also derive the growth rates of perturbations
in both Jordan and Einstein frames for the models in which 
the so-called $\phi$-matter dominated epoch \cite{Ame} 
is present.

\section{Background equations}

We start with the following 4-dimensional action
\begin{eqnarray}
\label{action}
S=\int {\rm d}^4 x \sqrt{-g} \left[
\frac{1}{2}f(R, \phi, X)+{\cal L}_m \right]\,,
\end{eqnarray}
where $g$ is a determinant of a metric $g_{\mu \nu}$,
$f$ is a function in terms of a Ricci scalar $R$, 
a scalar field $\phi$ and a kinetic term 
$X =-\phi^{,c} \phi_{,c}/2$. 
${\cal L}_m$ is a Lagrangian density  
for a pressureless matter 
whose energy density is given by $\rho_m$.
We use the metric signature $(-,+,+,+)$.

The gravitational field equation and the equation of motion 
of the field $\phi$ are given by 
\begin{eqnarray}
& & FG_{\mu \nu}=\frac12 (f-RF) g_{\mu \nu}+
F_{,\mu;\nu}-\kern1pt\vbox{\hrule height
1.2pt\hbox{\vrule width 1.2pt\hskip 3pt
\vbox{\vskip 6pt}\hskip 3pt\vrule width 0.6pt}\hrule
height 0.6pt}\kern1pt F g_{\mu \nu}\nonumber \\
& &~~~~~~~~~~+\frac12 f_{,X}
\phi_{,\mu}\phi_{,\nu}+T_{\mu \nu}^{(m)}\,, \\
& & (f_{,X}\phi^{,c})_{;c}+f_{,\phi}=0\,,
\end{eqnarray}
where $F=\partial f/\partial R$, $G_{\mu \nu}$ is an Einstein tensor, 
and $T_{\mu \nu}^{(m)}$ is an energy-momentum tensor
of the pressureless matter.

In a flat Friedmann-Robertson-Walker (FRW) metric
with a scale factor $a$, we obtain the following 
background equations
\begin{eqnarray}
\label{be1}
& & 3FH^2=f_{,X}X+\frac12 (FR-f)-3H\dot{F}+\rho_m\,,\\
\label{be2}
& & -2F\dot{H}=f_{,X}X+\ddot{F}-H\dot{F}+\rho_m\,,\\
\label{be3}
& & \frac{1}{a^3} \left( a^3 \dot{\phi} f_{,X} \right)^{\cdot}
-f_{,\phi}=0\,, \\
\label{be4}
& & \dot{\rho}_m+3H\rho_m=0\,,
\end{eqnarray}
where $H \equiv \dot{a}/a$, $R=6(2H^2+\dot{H})$, and 
a dot represents a derivative with respect to cosmic time $t$.

In order to confront the DE equation of state with 
observations such as SNIa, 
we rewrite Eqs.~(\ref{be1}) and (\ref{be2}) as follows:
\begin{eqnarray}
\label{mba1}
& & 3F_0 H^2=\rho_{\rde}+\rho_m\,, \\
\label{mba2}
& & -2F_0 \dot{H}=\rho_{\rde}+p_{\rde}+\rho_m\,,
\end{eqnarray}
where 
\begin{eqnarray}
& & \rho_{\rde}=\frac12 (FR-f)-3H\dot{F}+f_{,X}X
+3H^2(F_0-F), \\
& & p_{\rde}=\ddot{F}+2H\dot{F}-\frac12 (FR-f)
-(2\dot{H}+3H^2)(F_0-F). \nonumber \\
\end{eqnarray}
Here the subscript ``0'' represents present values.
It is easy to show that $\rho_{\rde}$ and $p_{\rde}$
defined in this way satisfy the usual energy 
conservation equation 
\begin{eqnarray}
\dot{\rho}_{\rde}+3H(\rho_{\rde}+p_{\rde})=0\,,
\end{eqnarray}
where we used Eq.~(\ref{be3}).
This was already shown to hold in the context of the scalar-tensor 
gravity \cite{Esp,GPRS} as well as the $f(R)$ gravity \cite{AGPT}.
We define the DE equation of state as 
\begin{eqnarray}
w_{\rde} &\equiv& \frac{p_{\rde}}{\rho_{\rde}} \nonumber \\
&=& -1+\frac{2f_{,X}X+2\ddot{F}-4H\dot{F}
-4\dot{H}(F_0-F)}{2f_{,X}X+FR-f-6H\dot{F}+
6H^2(F_0-F)}. \nonumber \\
\label{wde}
\end{eqnarray}

{}Integrating Eq.~(\ref{be4}) gives 
\begin{eqnarray}
\rho_m=3F_0 \Omega_m^{(0)} H_0^2
(1+z)^3\,,
\end{eqnarray}
where $z=a_0/a-1$ is a redshift and 
$\Omega_m^{(0)}$ is a present energy fraction of 
the non-relativistic matter.
Then by using Eqs.~(\ref{mba1}) and (\ref{mba2})
the equation of state $w_{\rde}$ can be 
expressed as
\begin{eqnarray}
w_{\rde}=-\frac{3r-(1+z)(\rd r/\rd z)}
{3r-3\Omega_m^{(0)}(1+z)^3}\,,
\end{eqnarray}
where $r=H^2(z)/H_0^2$.
This is the same relation as the one derived 
in Einstein gravity \cite{CST}.
Thus $w_{\rde}$ is constrained in the usual way 
from SNIa observations.
{}From Eq.~(\ref{wde}) we find that the evolution 
of $w_{\rm DE}$ depends upon the models of dark energy.
Hence one can test the viability of the models 
by confronting $w_{\rde}$ with observations.

If the scalar field $\phi$ is minimally coupled gravity, 
e.g., $f=R/8\pi G+2p(\phi, X)$, the structure of the 
Lagrangian density $p(\phi, X)$ can be reconstructed
by the evolution of the Hubble parameter 
$H(z)$ \cite{Tsure}.
For the models where the field $\phi$
is coupled to gravity or the models in which the Lagrangian 
includes non-linear terms in $R$, we need 
additional information to determine the strength of 
gravitational couplings.
This can be provided by considering the evolution of 
matter density perturbations. 

\section{Perturbation equations}

We consider the following perturbed metric with 
scalar metric perturbations $\Phi$ and $\Psi$
in a longitudinal gauge:
\begin{eqnarray}
\rd s^2 &=& - (1+2\Phi)\rd t^2 
+a^2 (1-2\Psi)\delta_{ij}\rd x^i \rd x^j\,.
\end{eqnarray}
We decompose the field into the background and inhomogeneous
parts: $\phi=\tilde{\phi}(t)+\delta \phi (t, {\bf x})$.
In what follows we drop the tilde for simplicity.
The energy momentum tensors of the non-relativistic matter
are decomposed as $T^{0}_{0}=-(\rho_m+\delta \rho_m)$
and $T^0_{\alpha}=-\rho_m v_{m, \alpha}$, where 
$v_m$ is a velocity potential.

The Fourier transformed perturbation equations 
are given by \cite{Hwang05}
\begin{widetext} 
\begin{eqnarray}
\label{per1}
& &3H(\dot{\Psi}+H\Phi)+\frac{k^2}{a^2}\Psi +\frac{1}{2F}
\biggl[ -\frac12 (f_{,\phi}\delta \phi+f_{,X}\delta X )+
\frac12 \dot{\phi}^2 (f_{,X \phi} \delta \phi +f_{,XX}\delta X)
+f_{,X}\dot{\phi}\delta \dot{\phi} 
-3H\delta \dot{F} \nonumber \\
& &+\left(3H^2+3\dot{H}-\frac{k^2}{a^2}
\right) \delta F+3\dot{F} (\dot{\Psi}+H\Phi)+
(3H\dot{F}-f_{,X}\dot{\phi}^2)\Phi
+\delta \rho_m \biggr]=0\,,\\
\label{per2}
& & f_{,X} \left[ \delta \ddot{\phi}+\left( 3H+
\frac{\dot{f}_{,X}}{f_{,X}} \right) \delta \dot{\phi}
+\frac{k^2}{a^2}\delta \phi -\dot{\phi} 
(3\dot{\Psi}+\dot{\Phi}) \right]
-2f_{,\phi}\Phi+\frac{1}{a^3} (a^3 \dot{\phi}
\delta f_{,X})^{\cdot}-\delta f_{,\phi}=0\,,\\
\label{per3}
& &\Psi=\Phi+\frac{\delta F}{F}\,,\\
\label{per4}
& & \delta \dot{\rho}_m+3H\delta \rho_m
=\rho_m \left( 3\dot{\Psi}-\frac{k^2}{a}v_m \right) 
\,, \\
\label{per5}
& & \dot{v}_m+Hv_m=\frac{1}{a}\Phi\,,
\end{eqnarray}
\end{widetext}  
where $k$ is a comoving wavenumber.

We define the gauge-invariant matter density perturbation
$\delta_m$, as 
\begin{eqnarray}
\delta_m \equiv \frac{\delta \rho_m}{\rho_m}
+3Hv\,,\quad {\rm where}
\quad
v \equiv a v_m\,.
\end{eqnarray}
Then Eqs.~(\ref{per4}) and (\ref{per5}) yield
\begin{eqnarray}
& & \dot{\delta}_m=-\frac{k^2}{a^2}v
+3(\Psi+Hv)^{\cdot}\,,\\
& & \dot{v}=\Phi\,,
\end{eqnarray}
from which we obtain
\begin{eqnarray}
\label{delright}
\ddot{\delta}_m+2H\dot{\delta}_m+\frac{k^2}{a^2}\Phi
=3\ddot{B}+6H\dot{B}\,,
\end{eqnarray}
where $B \equiv \Psi+Hv$.

Following the approach in Ref.~\cite{Boi,Esp,CST},  
we use a sub-horizon approximation under which the 
leading terms correspond to those containing  
$k^2$ and $\delta_m$ (or $\delta \rho_m$)
in Eq.~(\ref{delright}) and also in Eqs.~(\ref{per1})-(\ref{per2}).
Basically the terms on the r.h.s. of Eq.~(\ref{delright}) give
the contribution of the order $H^2 \Psi$, which implies that they 
are negligible relative to the term $(k^2/a^2)\Phi$ for the modes
deep inside the Hubble radius ($k^2 \gg a^2H^2$).

If the mass $m_\phi$ of the field perturbation $\delta \phi$ 
is larger than the term $k/a$ then we need to take into 
account this mass term. 
The expression of $m_\phi$ was derived in Ref.~\cite{Ameper} 
in coupled dark energy models with the Lagrangian density $p(\phi, X)$.
In Einstein gravity with a standard scalar field 
the mass squared is given by 
$m_\phi^2=V_{,\phi \phi}-2\dot{\phi}^2$.
When the field $\phi$ is responsible for dark energy 
the terms $V_{,\phi \phi}$ and $2\dot{\phi}^2$ are of 
order $H^2$ or less, which then gives $|m_\phi| \lesssim H$.
Hence the approximation neglecting the mass term $m_\phi$
relative to $k/a$ is justified in such a model.
There may be some specific $f(R,\phi,X)$ models in which 
the condition $|m_\phi| \ll k/a$ is violated, but we do not 
consider such cases.

Then Eq.~(\ref{delright}) is approximately given by 
\begin{eqnarray}
\label{delmeq}
\ddot{\delta}_m+2H\dot{\delta}_m+
\frac{k^2}{a^2}\Phi \simeq 0\,.
\end{eqnarray}
The next step is to express $\Phi$ in terms of $\delta_m$.
{}From Eq.~(\ref{per1}) we find 
\begin{eqnarray}
\label{Psi}
\frac{k^2}{a^2}\Psi \simeq \frac{1}{2F}
\left( \frac{k^2}{a^2} \delta F -\delta \rho_m \right)\,.
\end{eqnarray}
Eliminating the term $\Psi$ by using Eq.~(\ref{per3}) gives
\begin{eqnarray}
\label{Phi}
\frac{k^2}{a^2}\Phi \simeq
-\frac{k^2}{2a^2} \frac{\delta F}{F} 
-\frac{1}{2F} \delta \rho_m\,.
\end{eqnarray}

In what follows we shall study the case in which $F$ depends on 
$\phi$ and $R$ but not on $X$, i.e., 
\begin{eqnarray}
F=F(\phi, R)\,.
\end{eqnarray}
This actually includes most of dark energy models 
proposed in the current literature.
Then $\delta F$ in Eq.~(\ref{Phi}) is given by 
\begin{eqnarray}
\label{delF}
\delta F=F_{,\phi}\delta \phi +F_{,R} \delta R\,,
\end{eqnarray}
where $\delta R$ is 
\begin{eqnarray}
\label{delR}
\delta R &=& 2 \biggl[ -3 (\ddot{\Psi}+4H\dot{\Psi}
+H\dot{\Phi}+\dot{H}\Phi+4H^2\Phi) \nonumber \\
& &~~+\left( \frac{k^2}{a^2}-3\dot{H} \right) \Phi
-2\frac{k^2}{a^2}\Psi \biggr] \nonumber \\
&\simeq& -2\frac{k^2}{a^2} 
\left( \Phi+2\frac{\delta F}{F} \right)\,.
\end{eqnarray}
Again we used the fact that the first five terms 
in Eq.~(\ref{delR}) are of order 
$H^2 \Phi$, $H^2 \Psi$ or less.
Plugging Eq.~(\ref{delF}) into Eq.~(\ref{delR}), 
we find 
\begin{eqnarray}
\label{delR2}
\delta R \simeq -\frac{2k^2}{a^2} 
\frac{\Phi+\frac{2F_{,\phi}}{F}\delta \phi}
{1+4\frac{k^2}{a^2}\frac{F_{,R}}{F}}\,.
\end{eqnarray}

Taking notice that $\delta R$ includes the term $k^2/a^2$, 
the quantity $\delta f$ is approximately given by 
\begin{eqnarray}
\delta f&=&f_{,\phi}\delta \phi+f_{,X}\delta X
+f_{,R}\delta R \nonumber \\
&\simeq& F \delta R\,.
\end{eqnarray}
Then from Eq.~(\ref{per2}) we find
\begin{eqnarray}
f_{,X}\frac{k^2}{a^2} \delta \phi-F_{,\phi}
\delta R \simeq 0\,,
\end{eqnarray}
which leads to the following relation 
\begin{eqnarray}
\label{dp}
\delta \phi \simeq -2\frac{F_{,\phi}}
{f_{,X}\left( 1+4\frac{k^2}{a^2}\frac{F_{,R}}{F}\right)
+\frac{4F_{,\phi}^2}{F}}\Phi\,.
\end{eqnarray}
Plugging this into Eq.~(\ref{delR}), we get
\begin{eqnarray}
\label{dR}
\delta R \simeq -\frac{2k^2}{a^2}
\frac{f_{,X}}{f_{,X}\left( 1+4\frac{k^2}{a^2}\frac{F_{,R}}{F}\right)
+\frac{4F_{,\phi}^2}{F}}\Phi\,.
\end{eqnarray}
{}From Eqs.~(\ref{Phi}), (\ref{delF}), (\ref{dp}) and (\ref{dR})
the gravitational potential $\Phi$ is expressed as
\begin{eqnarray}
\label{Phieq}
\frac{k^2}{a^2}\Phi \simeq -\frac{\rho_m}{2F}
\frac{f_{,X}+4\left(f_{,X}\frac{k^2}{a^2}\frac{F_{,R}}{F}+
\frac{F_{,\phi}^2}{F}\right)}
{f_{,X}+3\left(f_{,X}\frac{k^2}{a^2}\frac{F_{,R}}{F}+
\frac{F_{,\phi}^2}{F}\right)}\delta_m\,,
\end{eqnarray}
where we used $\delta_m \simeq \delta \rho_m/\rho_m$
under the sub-horizon approximation.
Hence the equation (\ref{delmeq}) of matter 
perturbations yields
\begin{eqnarray}
\label{delful}
\ddot{\delta}_m+2H\dot{\delta}_m
-4\pi G_{\rm eff}\rho_m \delta_m \simeq 0\,,
\end{eqnarray}
where the effective gravitational ``constant'' on sub-horizon scales 
is given by 
\begin{eqnarray}
\label{Geff}
G_{\rm eff} \simeq \frac{1}{8\pi F}
\frac{f_{,X}+4\left(f_{,X}\frac{k^2}{a^2}\frac{F_{,R}}{F}+
\frac{F_{,\phi}^2}{F}\right)}
{f_{,X}+3\left(f_{,X}\frac{k^2}{a^2}\frac{F_{,R}}{F}+
\frac{F_{,\phi}^2}{F}\right)}\,.
\end{eqnarray}
{}From Eq.~(\ref{Phieq}) the gravitational potential is
\begin{eqnarray}
\label{gra}
\Phi \simeq -4\pi G_{\rm eff}
\frac{a^2}{k^2}\rho_m \delta_m\,,
\end{eqnarray}
which corresponds to a Poisson equation in the Fourier space.
In what follows we use the standard equality rather than the 
approximate equality ($\simeq$)
for the results obtained under the sub-horizon approximation.

We also define a parameter $\eta$
that characterizes the strength of an anisotropic stress:
\begin{eqnarray}
\label{Phidef}
\eta \equiv \frac{\Phi-\Psi}{\Psi}\,.
\end{eqnarray}
Using Eqs.~(\ref{per3}), (\ref{dp}) and (\ref{dR}) we obtain
\begin{eqnarray}
\label{etak}
\eta=\frac{2f_{,X}\frac{k^2}{a^2}\frac{F_{,R}}{F}
+\frac{2F_{,\phi}^2}{F}}
{f_{,X} \left(1+\frac{2k^2}{a^2}\frac{F_{,R}}{F}\right)+
\frac{2F_{,\phi}^2}{F}}\,.
\end{eqnarray}
The gravitational potential $\Psi$ satisfies 
\begin{eqnarray}
\label{Psieq}
\frac{k^2}{a^2}\Psi=-\frac{\rho_m}{2F}
\frac{f_{,X}+2\left(f_{,X}\frac{k^2}{a^2}\frac{F_{,R}}{F}+
\frac{F_{,\phi}^2}{F}\right)}
{f_{,X}+3\left(f_{,X}\frac{k^2}{a^2}\frac{F_{,R}}{F}+
\frac{F_{,\phi}^2}{F}\right)}\delta_m\,.
\end{eqnarray}
We define another parameter $q$ via
$(k^2/a^2)\Psi=-4\pi G_0 q \rho_m \delta_m$, where 
$G_0$ is a gravitational 
constant measured in the solar system experiments today.
Then $q$ is given by 
\begin{eqnarray}
q=\frac{1}{8\pi F G_0}
\frac{f_{,X}+2\left(f_{,X}\frac{k^2}{a^2}\frac{F_{,R}}{F}+
\frac{F_{,\phi}^2}{F}\right)}
{f_{,X}+3\left(f_{,X}\frac{k^2}{a^2}\frac{F_{,R}}{F}+
\frac{F_{,\phi}^2}{F}\right)}\,.
\end{eqnarray}

Defining a combination of parameters, $\Sigma=q(1+\eta/2)$, 
we obtain  
\begin{eqnarray}
\label{sig}
\Sigma=\frac{1}{8\pi F G_0}\,.
\end{eqnarray}
This agrees with the result in Ref.~\cite{AKS}
derived in the specific scalar-tensor model: 
$f=F(\phi)R+2X-2V(\phi)$\footnote{In Ref.~\cite{AKS}
the authors used a dimensionless function $\bar{F}=8\pi G F$, 
where $G$ is a bare gravitational constant. Then one has 
$\Sigma=G/\bar{F}G_0$.}.
For this model the parameter $\eta$ reduces
to $\eta=F_{,\phi}^2/(F+F_{,\phi}^2)$, which again 
agrees with the result given in Ref.~\cite{AKS}.

In order to confront the modified gravity models 
with the observations of weak lensing, 
we use the fact that the potential that characterizes 
the deviation of light rays corresponds to 
$\Phi_{\rm WL} \equiv \Phi+\Psi$ \cite{Schimd}.
{}From Eqs.~(\ref{Phieq}), (\ref{Psieq}) and (\ref{sig}) 
we find that the lensing potential satisfies
\begin{eqnarray}
\label{wlpotential}
\Phi_{\rm WL} \simeq -8\pi G_0
\frac{a^2}{k^2}\rho_m \delta_m \Sigma\,.
\end{eqnarray}
The effect of modified gravity theories manifests themselves 
in weak lensing in at least two ways. One is the multiplication of the 
term $\Sigma$ on the r.h.s. of Eq.~(\ref{wlpotential}).
Another is the modification of the evolution of $\delta_m$
due to the change of the effective gravitational constant 
$G_{\rm eff}$. The growth index $\gamma$
of matter perturbations is linked to the parameters
$\Sigma$ and $\eta$ \cite{AKS}.
Thus two parameters $(\Sigma, \eta)$ will be useful 
to detect the signature of modified gravity theories
from the future survey of weak lensing.

It can happen that the scales of weak lensing are in the 
region of non-linear clustering, in which case we need to 
map the linear power spectrum of the lensing potential  
into a non-linear one. In the context of modified gravity theories 
mapping formulas have not been well known.
We leave the analysis of such non-linear regimes in 
weak lensing for future work.

In Einstein gravity with the Lagrangian density 
$f=R/8\pi G +2p(\phi, X)$ we obtain 
the standard equation for matter perturbations:
\begin{eqnarray}
\label{delm}
\delta_m''+\left( \frac12 -\frac32 w_{\rm eff} \right)
\delta_m'-\frac32 \Omega_m \delta_m=0\,,
\end{eqnarray}
where a prime represents a derivative with respect to 
$N={\rm ln}\,a$, and 
\begin{eqnarray}
\label{wOme}
w_{\rm eff}=-1-\frac{2}{3}\frac{H'}{H}\,, \quad
\Omega_m=\frac{\rho_m}{3FH^2}\,.
\end{eqnarray}
If $w_{\rm eff}$ and $\Omega_m$ are constants then 
the solution for Eq.~(\ref{delm}) is 
\begin{eqnarray}
\label{delmso}
\delta_m=c_+ a^{n_+}+c_- a^{n-}\,,
\end{eqnarray}
where 
\begin{eqnarray}
n_{\pm}=\frac14 \left[ 3w_{\rm eff}-1+
\sqrt{(3w_{\rm eff}-1)^2+24\Omega_m}
\right]\,.
\end{eqnarray}
One has $w_{\rm eff} \simeq 0$ and 
$\Omega_m \simeq 1$ during a matter-dominated 
epoch provided that the contribution of the scalar field is negligible.
Hence the matter perturbation grows as 
$\delta_m \propto a \propto t^{2/3}$.
However the evolution of $\delta_m$ is modified once 
the energy density of the scalar field becomes important 
relative to the matter density.

At the end of this section we consider the Brans-Dicke 
theory \cite{BD}:
\begin{eqnarray}
f(R,\phi,X)=\frac{\phi}{8\pi}R+\frac{\omega_{\rm BD}}
{4\pi \phi}X\,,
\end{eqnarray}
where $\omega_{\rm BD}$ is a Brans-Dicke parameter.
In this case the effective gravitational constant is given by 
\begin{eqnarray}
\label{GBD}
G_{\rm eff}=\frac{1}{\phi}
\frac{4+2\omega_{\rm BD}}
{3+2\omega_{\rm BD}}\,.
\end{eqnarray}
which agrees with the result 
in Ref.~\cite{Damour}.
We have $G_{\rm eff} \to 1/\phi$ in the 
General Relativity (GR) limit ($\omega_{\rm BD} \to \infty$).
The deviation from GR is significant when $\omega_{\rm BD}$
is not much larger than unity.
When $\omega_{\rm BD}=0$, for example,  the effective
gravitational constant is $4/3$ times larger than that 
in the GR case. This modifies the evolution of matter 
perturbations.
However local gravity experiments place the bound on 
the present value of the Brans-Dicke parameter
as $\omega_{\rm BD, 0}>4 \times 10^4$ \cite{BDcon}.
This shows that unless $\omega_{\rm BD}$ is 
very much smaller than the present value during the matter epoch 
it is difficult to see the signature of modified gravity
in the large-scale structure formation.

\section{~$f(R)$ gravity}

In this section we study the evolution of matter perturbations
in modified gravity theories where $f$ is the function of $R$
only. In this case the effective gravitational constant is given by 
\begin{eqnarray}
G_{\rm eff}=\frac{1}{8\pi F}
\frac{1+4\frac{k^2}{a^2R}m}{1+3\frac{k^2}{a^2R}m}\,,
\end{eqnarray}
where
\begin{eqnarray}
m=\frac{RF_{,R}}{F}\,.
\end{eqnarray}
The parameter $m$ was first introduced in 
Ref.~\cite{AGPT}.  This characterizes the deviation from 
the $\Lambda$CDM model ($f(R)=R/8\pi G-\Lambda$).
The anisotropic parameter $\eta$ is given by 
\begin{eqnarray}
\eta=\frac{2\frac{k^2}{a^2 R}m}
{1+2\frac{k^2}{a^2R}m}\,.
\end{eqnarray}

In what follows we shall consider two different situations:
(i) $\frac{k^2}{a^2R}m \gg 1$ and 
(ii) $\frac{k^2}{a^2R}m \ll 1$.

\subsection{$\frac{k^2}{a^2R}m \gg 1$}

In this case one has $G_{\rm eff} \simeq \frac{1}{8\pi F}\frac43$, 
which thus corresponds to Brans-Dicke theory with 
$\omega_{\rm BD}=0$ \cite{Chiba}, see Eq.~(\ref{GBD}).
Note that the anisotropic parameter is of order unity 
($\eta \simeq 1$) in such a case.
Since the condition $k^2/a^2R \gg 1$ holds 
under the sub-horizon approximation ($k \gg aH$), 
one can in fact realize $\frac{k^2}{a^2R}m \gg 1$
provided that $m$ is not very much smaller than unity.
Then from Eq.~(\ref{delm}) the matter 
perturbation equation is approximately given by 
\begin{eqnarray}
\label{delm2}
\delta_m''+\left( \frac12 -\frac32 w_{\rm eff} \right)
\delta_m' - 2\Omega_m \delta_m=0\,,
\end{eqnarray}
where $w_{\rm eff}$ and $\Omega_m$ are 
defined in Eq.~(\ref{wOme}).

In order to estimate the evolution of $\delta_m$ and 
$\Phi$ analytically, let us consider the constant 
$m$ model. i.e., 
\begin{eqnarray}
f(R)=\alpha R^{1+m}-\Lambda\,,
\end{eqnarray}
where $\alpha$ and $\Lambda$ are constants.
In Ref.~\cite{AGPT} it was shown that the 
matter-dominated epoch corresponds to 
a fixed point ``$P_5$'' satisfying 
\begin{eqnarray}
\label{P5}
w_{\rm eff}=-\frac{m}{1+m}\,, \quad
\Omega_m=1-\frac{m(7+10m)}
{2(1+m)^2}\,,
\end{eqnarray}
where $|m| \ll 1$.

Plugging Eq.~(\ref{P5}) into Eq.~(\ref{delm2}), 
we obtain the solution for $\delta_m$ in the form 
(\ref{delmso}) with  
\begin{eqnarray}
\label{npm}
n_{\pm}=\frac{-(1+4m) \pm 
\sqrt{(3+4m)(11-28m)}}{4(1+m)}\,.
\end{eqnarray}
Since the growing mode corresponds to the power-law
index $n_+$, the matter perturbation evolves as 
\begin{eqnarray}
\label{an}
\delta_m \propto a^{n_+} \propto t^{\tilde{n}_+}\,,
\end{eqnarray}
where 
\begin{eqnarray}
\tilde{n}_+=
\frac{\sqrt{(3+4m)(11-28m)}-1-4m}{6}\,.
\end{eqnarray}

In Ref.~\cite{AGPT} it was found that the viable 
matter epoch exists only for positive $m$
close to 0. 
In the case of negative $m$ the matter point $P_5$
is unstable against perturbations around the 
fixed point.
When $m=-1/4$ one has $w_{\rm eff}=1/3$ and 
$\Omega_m=2$, which
corresponds to a $\phi$ matter-dominated 
epoch ($\phi$MDE) \cite{APT}.
For the models $f(R)=R-\beta/R^n$ ($n>0$) it was shown 
in Ref.~\cite{APT} that the standard matter era is 
replaced by the $\phi$MDE. 
{}From Eq.\,(\ref{npm}) we find that the matter perturbation 
evolves as $\delta_m \propto a^2 \propto t$, which grows 
more rapidly than in the standard case 
($\delta_m \propto a \propto t^{2/3}$).

The power-law indices $n_+$ and $\tilde{n}_+$ are positive for 
$0<m<(\sqrt{73}-3)/16$, whereas they are negative for 
$(\sqrt{73}-3)/16<m<11/28$. When $m>11/28$ the matter 
perturbation exhibits a damped oscillation.
Both $n_+$ and $\tilde{n}_+$ get larger as $m$ decreases to zero.
However we have to caution that we can not take the limit 
$m \to 0$ because of the breakdown of the condition 
$\frac{k^2}{a^2R}m \gg 1$.
In this limit the evolution of $\delta_m$ is no longer described 
by the solution (\ref{an}).

{}From Eq.~(\ref{gra}) we find that the gravitational 
potential evolves as 
\begin{eqnarray}
\label{Phi2}
\Phi \propto t^{p_+}\,,~~
p_+=\frac{\sqrt{(3+4m)(11-28m)}+4m-5}{6}\,.
\end{eqnarray}
One has $p_+>0$ for $0<m<1/4$ and $p_+<0$
for $1/4<m<11/28$. Hence $\Phi$ is not constant 
except for the special case $m=1/4$ and the 
$\phi$MDE case $m=-1/4$.
The variation of the gravitational potential leads to 
an Integrated-Sachs-Wolfe effect in the CMB spectrum.
Thus it should be possible to constrain the magnitude 
of $m$ from CMB observations.

\subsection{$\frac{k^2}{a^2R}m \ll 1$}

Let us next consider the case in which the condition, 
$\frac{k^2}{a^2R}m \ll 1$, is satisfied 
on the scales around which large-scale structure is formed.
In this case one has $G_{\rm eff} \simeq 
\frac{1}{8\pi F} \left(1+\frac{k^2}{a^2R}m \right)$ and 
$\eta \simeq 2\frac{k^2}{a^2R}m$.
The matter perturbation equation is 
approximately given by 
\begin{eqnarray}
\delta_m''+\frac12 \delta_m' -\frac32 
\left(1 +\frac{k^2}{a^2R} m \right)\delta_m=0\,.
\end{eqnarray}
Since the condition $m \ll \frac{k^2}{a^2R}m \ll 1$
holds under the sub-horizon approximation, 
we only pick up the correction terms that contain 
$\frac{k^2}{a^2R}m$.

Expressing the solutions of this equation in the form 
$\delta_m=\exp(\int \omega\,{\rm d}N)$ and using 
the approximation $|\omega'| \ll \omega^2$, we obtain 
the growing-mode solution
\begin{eqnarray}
\omega_+=1+\frac35 \frac{k^2}{a^2 R}m\,.
\end{eqnarray}
If $m$ is constant, the second term on the r.h.s. of this 
equation is proportional to $a=e^N$ during the matter era.
Then the evolution of the matter perturbation 
is given by 
\begin{eqnarray}
\label{delm3}
\delta_m \propto a^{1+\frac35 \beta} 
\propto t^{\frac23 \left(1+\frac35 \beta \right)}\,,
\end{eqnarray}
where 
\begin{eqnarray}
\beta \equiv \frac{k^2}{a^2 R N}m
=C k^2 m\,
\frac{a}{{\rm ln}\,a}\,.
\end{eqnarray}
Here we have introduced a constant $C$ satisfying 
the relation $1/(a^2R)=Ce^N$.
The gravitational potential evolves as
\begin{eqnarray}
\label{Phi3}
\Phi \propto \left( 1+\beta
\right) t^{\frac25 \beta}\,.
\end{eqnarray}

In the limit $\beta \to 0$ one obtains the standard 
result: $\delta_m \propto t^{2/3}$ and 
$\Phi={\rm constant}$.
When $\beta$ is positive, the growth rates of $\delta_m$
and $\Phi_m$ are larger than in the standard case.
If $\beta$ grows to the order of unity, 
the results (\ref{delm3}) and (\ref{Phi3}) are no longer valid. 

In order to satisfy the local gravity constraint (LGC), 
we require that the condition $\frac{k^2}{a^2R}m \ll 1$
holds at the present epoch.
The severest constraint may be obtained by laboratory
experiments in which a strong modification of gravity 
is not observed on the scales up to 
$\lambda_k \sim a/k \sim $\,1 mm.
This gives the following constraint
\begin{eqnarray}
\label{mcon}
m(z=0) \ll (\lambda_k/H_0^{-1})^2 \sim 10^{-58}\,,
\end{eqnarray}
where we used $R \sim H_0^2$ and
$H_0^{-1} \sim 10^{28}$\,cm. 
Note that this agrees with the result in Ref.~\cite{AT} that was 
derived by using the effective mass of a scalar-field 
potential in the Einstein frame (see Refs.~\cite{LGCrecent}
for recent works of LGC in $f(R)$ gravity).
The condition (\ref{mcon}) can be weakened by taking 
into account the fact that the scalar curvature $R$ in the regime 
of a local structure such as an earth is much larger than
the cosmological one.
In solar system experiments the scale $\lambda_k$
corresponds to a value around $\lambda_k=1$\,au, 
which is much larger than in the case of laboratory 
experiments.
In such cases the constraint on $m$ becomes much weaker 
than the one given in Eq.~(\ref{mcon}), 
although it is not easy to obtain the values of $m$
close to the order of unity \cite{AT}.

The variable $m$ generally changes with time
apart from the model $f=\alpha R^{1+m}-\Lambda$
discussed above.
One can consider models in which $m$ satisfies 
$\frac{k^2}{a^2R}m \gg 1$ during the matter epoch
and then enters the regime $\frac{k^2}{a^2R}m \ll 1$
in the dark energy era with the decrease of $m$.
The anisotropic stress parameter $\eta$ decreases from 
$1$ to $2\frac{k^2}{a^2R}m$ together with the 
change of the quantity $\Sigma$ given 
in Eq.~(\ref{sig}).
It will be of interest to place observational constraints 
on such models by using the future data of 
weak lensing as well as the LSS data.

\section{Scalar-tensor gravity}

In this section we shall consider scalar-tensor gravity 
with a Lagrangian density 
\begin{eqnarray}
\label{scalag}
f(R,\phi,X)=F(\phi) R+2p(\phi, X)\,.
\end{eqnarray}
Note that this includes most of scalar-field dark energy 
models such as quintessence \cite{quin}, 
k-essence \cite{kes} and tachyons \cite{tach}.
In this case the effective gravitational constant $G_{\rm eff}$ 
and the anisotropic parameter $\eta$ are given by 
\begin{eqnarray}
G_{\rm eff}&=&\frac{1}{8\pi F}
\frac{2p_{,X}+4F_{,\phi}^2/F}{2p_{,X}+3F_{,\phi}^2/F}\,,\\
\eta &=& \frac{F_{,\phi}^2}{p_{,X}F+F_{,\phi}^2}\,.
\end{eqnarray}
In the case of quintessence with the Lagrangian density 
$p=X-V(\phi)$ the above results agree with those obtained by 
Boisseau {\it et al.} \cite{Boi} and by 
Amendola {\it et al.} \cite{AKS}.

One can study the evolution of perturbations in the Jordan frame
as we have done in $f(R)$ gravity models.
Alternatively the dynamics of perturbations can be 
understood by making a conformal transformation to 
the Einstein frame. This is particularly useful when we relate
scalar-tensor models with coupled DE scenarios \cite{Ame}
extensively studied by many authors
(see \cite{CST} for references).
We shall make a conformal transformation \cite{Maeda}
\begin{eqnarray}
\tilde{g}_{\mu \nu}=\Omega g_{\mu \nu}\,,\quad
\Omega=\sqrt{F}\,,
\end{eqnarray}
where a tilde denotes quantities in the Einstein frame.
Then the action in the Einstein frame is given by 
\begin{eqnarray}
\label{actione}
\t{S}=\int {\rm d}^4 \t{x} \sqrt{-\t{g}} \left[
\frac{1}{2\kappa^2}\tilde{R}+\tilde{p}(\phi, \t{X}) 
+\t{{\cal L}}_m (\phi) \right]\,,
\end{eqnarray}
where $\kappa^2=8\pi G$ 
($G$ is a bare gravitational constant) and 
\begin{eqnarray}
\label{tp}
\t{p}(\phi, \t{X})=\frac32 \left( \frac{F_{,\phi}}{F} 
\right)^2 \t{X} +\frac{1}{F^2}\,p(\phi, F\t{X})\,.
\end{eqnarray}
In what follows we shall use the unit $\kappa^2=1$, 
but we restore the gravitational constant $G$ 
when it is needed.

We also have the following relations
\begin{eqnarray}
\label{conformal}
\tilde{a}=\sqrt{F}a,~~{\rm d}\tilde{t}=\sqrt{F} {\rm d}t,~~
\tilde{\rho}_m=\rho_m/F^2.
\end{eqnarray}
Then the continuity equation (\ref{be4}) is transformed as 
\begin{eqnarray}
\frac{\rd}{\rd \tilde{t}}\tilde{\rho}_m
+3\tilde{H} \tilde{\rho}_m=
Q(\phi) \tilde{\rho}_m \frac{\rd \phi}
{\rd \tilde{t}}\,,
\end{eqnarray}
where 
\begin{eqnarray}
\label{Qdef}
Q(\phi)=-\frac{F_{,\phi}}{2F}\,.
\end{eqnarray}
Hence the non-relativistic matter is coupled to the field $\phi$
through the coupling $Q(\phi)$.

The perturbations in the Einstein frame are related to 
those in the Jordan frame via \cite{Hwang05}
\begin{eqnarray}
\tilde{\delta}_m=\delta_m-\frac{2\delta F}{F}\,,\quad
\tilde{\Phi}=\Phi+\frac{2\delta F}{F}\,.
\label{Phie}
\end{eqnarray}
Under the sub-horizon approximation 
one can regard $\tilde{\delta}_m \simeq \delta_m$, 
whereas the $2\delta F/F$ contribution can not be 
neglected for the gravitational potential.
Using the relation $\tilde{p}_{,\t{X}}=\frac32 
\left( \frac{F_{,\phi}}{F} \right)^2+\frac{p_{,X}}{F}$
together with the definition of $Q$ given in Eq.~(\ref{Qdef}), 
we find that Eq.~(\ref{delful}) is written in terms of the quantities 
in the Einstein frame:
\begin{eqnarray}
& &\frac{\rd^2}{\rd \tilde{t}^2} \tilde{\delta}_m+\tilde{H}
\left(2+Q\frac{1}{\t{H}} \frac{\rd \phi}{\rd \t{t}}
\right)\frac{\rd}{\rd \t{t}} \tilde{\delta}_m \nonumber \\
& &-4\pi G \left( 1+\frac{2Q^2}{\t{p}_{,\t{X}}}
\right) \tilde{\rho}_m \tilde{\delta}_m=0\,.
\label{delmQ}
\end{eqnarray}
In this section we use a prime to represent a derivative with respect to 
the number of $e$-foldings $\tilde{N}=\int \t{H}\,\rd \t{t}$
in the Einstein frame.
Then the above equation can be written as 
\begin{eqnarray}
\label{delQeq}
\t{\delta}_m''+\left( 2+ \frac{\t{H}'}{\t{H}}+Q\phi'
\right) \t{\delta}_m'-\frac32 \tilde{\Omega}_m
\left( 1+\frac{2Q^2}{\t{p}_{,\t{X}}}
\right) \t{\delta}_m=0, \nonumber \\
\end{eqnarray}
where $\t{\Omega}_m=8\pi G \t{\rho}_m/3\t{H}^2$.
This fully agrees with the result in coupled dark energy scenarios 
derived by Amendola \cite{Ameper} without any reference 
to the Jordan frame (see also Ref.~\cite{Lucaper}). 

Equation (\ref{delmQ}) shows that  the effective 
gravitational constant in the Einstein frame is given by 
\begin{eqnarray}
\label{GeffE}
\tilde{G}_{\rm eff}=G \left(1+\frac{2Q^2}
{\t{p}_{,\t{X}}} \right)
=G\,\frac{2p_{,X}+4F_{,\phi}^2/F}
{2p_{,X}+3F_{,\phi}^2/F}.
\end{eqnarray}
{}From Eqs.~(\ref{dp}) and (\ref{Phie}) the 
gravitational potential in the Einstein frame is 
\begin{eqnarray}
\label{tilPhi}
\tilde{\Phi}=\frac{2p_{,X}+3F_{,\phi}^2/F}
{2p_{,X}+4F_{,\phi}^2/F}\Phi\,, 
\end{eqnarray}
which satisfies the relation 
\begin{eqnarray}
\label{tilPhi2}
\tilde{\Phi}=-4\pi G \frac{\tilde{a}^2}{k^2}
\tilde{\rho}_m \tilde{\delta}_m\,.
\end{eqnarray}
The effective gravitational potential acting on the matter 
is not $\tilde{\Phi}$ but 
$\tilde{\Phi}^*=\tilde{\Phi}+Q\delta \phi=\Phi$ \cite{Ameper}, 
i.e., that in the Jordan frame.
In fact, from Eqs.~(\ref{GeffE}) and (\ref{tilPhi2}), we obtain 
\begin{eqnarray}
\label{tilPhi3}
\tilde{\Phi}^*=-4\pi \tilde{G}_{\rm eff}
\frac{\tilde{a}^2}{k^2}
\tilde{\rho}_m \tilde{\delta}_m\,.
\end{eqnarray}

In order to see the effect of an interaction between the field $\phi$
and the matter analytically it is convenient to study 
the constant $Q$ case, i.e., 
\begin{eqnarray}
F(\phi)=e^{-2Q\phi}\,.
\end{eqnarray}
In the case of an ordinary field with an exponential potential 
[$\t{p}(\phi, \t{X})=\t{X}-ce^{-\lambda \phi}$], it is known that 
there exists a $\phi$MDE scaling solution satisfying 
$\t{\Omega}_\phi=\t{w}_{\rm eff}
=2Q^2/3={\rm constant}$ \cite{Ame}.
More generally the existence of scaling solutions restricts the form 
of the Lagrangian density to be 
\begin{eqnarray}
\t{p}(\phi, \t{X})=\t{X}\,g(Y)\,,\quad
Y=\t{X}e^{\lambda \phi}\,,
\end{eqnarray}
where $g$ is an arbitrary function and $\lambda$ is 
a constant quantity \cite{Tsuji}.
It was further shown in Ref.~\cite{AQTW} that the $\phi$MDE
exists for the models of the type
\begin{eqnarray}
\label{gY}
g(Y)=c_0+\sum_{n>0} c_nY^{-n}\,,
\end{eqnarray}
where $c_0$ ($>0$) and $c_n$ are constants. 
Note that quintessence with an exponential potential 
corresponds to the case $g(Y)=1-c_1/Y$.
For the models (\ref{gY}) we have the following relations 
during the $\phi$MDE \cite{AQTW}:
\begin{eqnarray}
\tilde{\Omega}_\phi=\tilde{w}_{\rm eff}
=\frac{2Q^2}{3c_0},\quad \phi'=-\frac{2Q}{c_0},
\quad \t{p}_{,\t{X}}=c_0\,,
\label{rel}
\end{eqnarray}
where $\tilde{\Omega}_\phi$ is an energy fraction of the 
scalar field satisfying $\tilde{\Omega}_\phi+\tilde{\Omega}_m=1$.

Then from Eq.~(\ref{delQeq}), the solution for 
matter perturbations during the $\phi$MDE
is given by the form (\ref{delmso}) with the power-law indices
\begin{eqnarray}
\t{n}_+=1+\frac{2Q^2}{c_0}\,, \quad
\t{n}_-=-\frac32+\frac{Q^2}{c_0}\,.
\end{eqnarray}
Since the scale factor grows as $\tilde{a} \propto 
\t{t}^{\frac{2c_0}{3c_0+2Q^2}}$, the evolution of 
the matter perturbation in the Einstein frame is given by 
\begin{eqnarray}
\t{\delta}_m \propto  \t{a}^{1+\frac{2Q^2}{c_0}}
\propto \t{t}^{\frac{2c_0+4Q^2}{3c_0+2Q^2}}\,.
\end{eqnarray}
{}From Eq.~(\ref{rel}) we obtain 
\begin{eqnarray}
\phi=\phi_0-\frac{4Q}{3c_0+2Q^2} 
\,{\rm ln}\,\t{t}\,.
\end{eqnarray}
Using the relations (\ref{conformal}) between two frames we find 
\begin{eqnarray}
t \propto  \t{t}^{\frac{3c_0-2Q^2}{3c_0+2Q^2}}\,,\quad
a \propto \t{t}^{\frac{2c_0-4Q^2}{3c_0+2Q^2}}\,.
\end{eqnarray}
Hence the evolution of matter perturbations in the Jordan frame 
is given by 
\begin{eqnarray}
\label{Jo}
\delta_m \propto a^{\frac{2c_0+4Q^2}{2c_0-4Q^2}}
\propto t^{\frac{2c_0+4Q^2}{3c_0-2Q^2}}\,.
\end{eqnarray}
Thus, in the presence of the coupling $Q$, the growth rate of 
matter perturbations is larger than in the case of Einstein gravity.
{}From Eqs.~(\ref{tilPhi2}) and (\ref{tilPhi3}) we find that the 
gravitational potential is constant in both Jordan and Einstein frames:
\begin{eqnarray}
\Phi \propto t^0\,, \quad \t{\Phi} \propto \t{t}^0\,,
\end{eqnarray}
which is a rather peculiar property of the $\phi$MDE.
Recall that this property also holds for the $\phi$MDE
solution in the $f(R)$ gravity.

{}From Eq.~(\ref{tp}) the Lagrangian density 
in the Jordan frame corresponding to the $\phi$MDE solution 
is given by 
\begin{eqnarray}
f(R,\phi,X)&=&e^{-2Q \phi} \biggl[ R
+2(c_0-6Q^2)X \nonumber \\
&&+\sum_{n>0} 2c_n X^{1-n} e^{-n(\lambda+2Q)\phi}
\biggr]\,.
\end{eqnarray}
Thus the field $\phi$ has a universal coupling $e^{-2Q\phi}$.
For the model $g(Y)=1-c_1/Y$ the above Lagrangian 
density can be viewed as the dilaton gravity with an 
exponential potential $V(\phi)=2c_1e^{-(\lambda+2Q)\phi}$.
It is interesting that string theory can give rise to 
the $\phi$MDE solution along which $\Omega_m$ and 
$w_{\rm eff}$ are constants in both Jordan and 
Einstein frames.
We note that $Q$ is required to be smaller than the order 
of unity to reproduce a standard matter era \cite{Ame}, 
whereas string theory typically provides the coupling $Q$ of 
order one at the perturbative regime \cite{Gas}. 
In the region $\phi \gg 1$, however, 
the coupling may become weak as in the context of a runaway dilaton 
scenario \cite{runaway}.

\section{Conclusions}

In this paper we derived the matter perturbation equation (\ref{delful}) 
with the effective gravitational constant (\ref{Geff}) for a 
Lagrangian density $f(R,\phi,X)$ without a direct coupling between 
$R$ and $X$.
This analysis covers most of modified 
gravity models proposed in the current literature
and will be useful to detect the deviation from the 
$\Lambda$CDM model from the future surveys such as 
weak lensing and LSS. 
We have also evaluated the anisotropic parameter $\eta$
and the quantity $\Sigma=q(1+\eta/2)$ in order to
confront the models with future observations of weak lensing, 
see Eqs.~(\ref{etak}) and (\ref{sig}).
We have applied our results to (i) $f(R)$ gravity and 
(ii) scalar-tensor gravity with the Lagrangian density 
$f(R)=F(\phi)R+2p(\phi, X)$.

In $f(R)$ gravity models the effective gravitational constant
has a scale-dependent term $\frac{k^2}{a^2R}m$, where 
$m=Rf_{,RR}/f_{,R}$ characterizes the deviation of the $\Lambda$CDM 
model. The local gravity constraint is satisfied for the models in which 
the condition, $\frac{k^2}{a^2R}m \ll 1$, holds
for the scale of the order $a/k=1$\,mm at the present epoch.
If we take the cosmological value $R \sim H_0^2$, 
this gives a very stringent constraint: $m(z=0) \ll 10^{-58}$.
This is weakened by using a local value of $R$ much 
larger than $H_0^2$. 
One can consider models in which the condition, 
$\frac{k^2}{a^2R}m \gg 1$, holds during the matter epoch 
on the scales around which large-scale
structure is formed. For the constant $m$ model 
($f=\alpha R^{1+m}-\Lambda$) we have analytically derived the 
evolution of matter perturbations $\delta_m$ as well as 
gravitational potentials $\Phi$ during the matter-dominated epoch.
Even when $m \ll 1$ this is different from the evolution in Einstein gravity 
($\delta_m \propto a$ and $\Phi=$constant), which will be 
useful to place constraints on the value $m$ from future 
high-precision observations.

The scalar-tensor gravity with the Lagrangian density 
$f(R)=F(\phi)R+2p(\phi, X)$ correspond to coupled DE models 
in the Einstein frame with a coupling 
$Q(\phi)=-F_{,\phi}/2F$ between the scalar field and dark matter. 
We reproduced the equation of matter perturbations in coupled DE 
scenarios with the k-essence Lagrangian 
density $\t{p}(\phi, \t{X})$ by making 
a conformal transformation to the Einstein frame.
Since the evolution of perturbations in coupled DE models has been 
extensively studied in literature, it is convenient to pay attention 
to the relation between Jordan and Einstein frames
in order to discuss perturbations in the 
scalar-tensor gravity. In fact, for the models in which the $\phi$ matter 
dominated epoch ($\phi$MDE) is present, we analytically derived growth 
rates of perturbations in both Jordan and Einstein frames. 
We also obtained the form of the Lagrangian 
density in the Jordan frame giving rise to 
the $\phi$MDE solution.

The difference between $f(R)$ gravity and scalar-tensor gravity may 
be understood in the following way. 
Taking into account a mass $M$ of the perturbation in the 
field $\phi$, the effective gravitational constant in Eq.~(\ref{GeffE})
is given by \cite{Ameper}
\begin{eqnarray}
\label{GYukawa}
\tilde{G}_{\rm eff} \simeq 
G \left(1+\frac{2Q^2}{\tilde{p}_{,\tilde{X}}}
e^{-M \ell} \right)\,,
\end{eqnarray}
where $\ell$ is a length scale.
Here we assumed that the sound speed $c_s$ 
of the field $\phi$ is of order unity.
In scalar-tensor models, even if the mass $M$ is very light 
as $M \sim H_0$, the second term on the r.h.s.
of (\ref{GYukawa})
can be much smaller than unity to satisfy the local gravity constraints
by choosing a small coupling $Q^2=(F_\phi/2F)^2 \ll 1$ 
(provided that $\tilde{p}_{,\tilde{X}}$ is of order one).
In $f(R)$ modified gravity models, however, the coupling $Q$
is fixed as $Q=-1/\sqrt{6}$ \cite{APT}.
Hence we have to choose a heavy mass $M^2 \simeq 1/F_{,R}$ to 
satisfy the local gravity constraints, i.e., $e^{-M \ell} \ll 1$
for the scale $\ell \sim 1$\,mm \cite{AT}.
In fact this is equivalent to choosing very small values of 
$m$ given in Eq.~(\ref{mcon}).
In scalar-tensor gravity the coupling-dependent term is more 
important than the scale-dependent term provided that $M \ll k/a$, 
whereas in $f(R)$ gravity the scale-dependent term plays a crucial 
role because of the fixed large coupling $Q$.

Although the perturbation equation we derived can be applied 
to many modified gravity models, it does not cover the DE models
in which higher-order curvature corrections such as a Gauss-Bonnet (GB)
term are present \cite{GB0}. 
In Ref.~\cite{GB1} the equation of matter perturbations 
was derived in the presence of the GB term coupled to a 
scalar field $\phi$ to place constraints on GB DE models.
A recent paper \cite{GB2} shows that the GB energy fraction in the 
present universe is severely constrained by solar system tests.  
Moreover it is known that tensor perturbations typically show negative instabilities
if the GB term is responsible for the accelerated expansion \cite{instability}.
It will be of interest to extend our analysis to more general models
that include such higher-order curvature corrections.

\section*{ACKNOWLEDGMENTS}
This work is supported by JSPS
(Grant No.\,30318802).
I thank Luca Amendola for useful discussions and drawing 
my attention to several important points.
I am also grateful to Wayne Hu, Martin Kunz 
and David Polarski for useful comments.
I also thank Istvan Laszlo for pointing out a typo in 
Eq.~(\ref{Phidef}) that existed in the version 3 
of this paper.


\begin{thebibliography}{40}
  
\bibitem{SN} 
S.~Perlmutter {\it et al.},
Astrophys.\ J.\  {\bf 517}, 565 (1999); 
A.~G.~Riess {\it et al.},
Astron.\ J.\  {\bf 116}, 1009 (1998);
Astron.\ J.\  {\bf 117}, 707 (1999).

\bibitem{SN2} 
P.~Astier {\it et al.},
Astron.\ Astrophys.\  {\bf 447}, 31 (2006); 
A.~G.~Riess {\it et al.},
arXiv:astro-ph/0611572;
W.~M.~Wood-Vasey {\it et al.},
arXiv:astro-ph/0701041.
	
\bibitem{CMB}
D.~N.~Spergel {\it et al.}  [WMAP Collaboration],
Astrophys.\ J.\ Suppl.\ {\bf 148}, 175 (2003);
D.~N.~Spergel {\it et al.},
arXiv:astro-ph/0603449.	
	
\bibitem{BAO}
D.~J.~Eisenstein {\it et al.}  [SDSS Collaboration],
Astrophys.\ J.\  {\bf 633}, 560 (2005).
	
\bibitem{review} 
V.~Sahni and A.~A.~Starobinsky, 
Int.\ J.\ Mod.\ Phys.\ D \textbf{9}, 373 (2000); 
S.~M.~Carroll,
Living Rev.\ Rel.\  \textbf{4}, 1 (2001); 
V.~Sahni, 
Lect.\ Notes Phys.\  \textbf{653}, 141 (2004) {[}arXiv:astro-ph/0403324];
T.~Padmanabhan, 
Phys.\ Rept.\  \textbf{380}, 235 (2003); P.~J.~E.~Peebles and
B.~Ratra, 
Rev.\ Mod.\ Phys.\  \textbf{75}, 559 (2003); 
S.~Nojiri and S.~D.~Odintsov, Int.\ J.\ Geom.\ Meth.\ Mod.\ Phys.\ 
\textbf{4}, 115 (2007).	
	
\bibitem{CST}		
E.~J.~Copeland, M.~Sami and S.~Tsujikawa,
Int.\ J.\ Mod.\ Phys.\  D {\bf 15}, 1753 (2006).		
	
\bibitem{quin} 
Y.~Fujii, Phys.\ Rev.\ D \textbf{26}, 2580 (1982);
C.~Wetterich, Nucl. \ Phys\ B. \textbf{302}, 668 (1988); B. Ratra
and J. Peebles, Phys. \ Rev\ D \textbf{37}, 321 (1988); 
T.~Chiba, N.~Sugiyama and T.~Nakamura,
Mon.\ Not.\ Roy.\ Astron.\ Soc.\  {\bf 289}, L5 (1997);
I.~Zlatev,
L.~M.~Wang and P.~J.~Steinhardt, Phys.\ Rev.\ Lett.\  \textbf{82},
896 (1999).		

\bibitem{kes} 
T.~Chiba, T.~Okabe and M.~Yamaguchi,
Phys.\ Rev.\ D {\bf 62}, 023511 (2000);
C.~Armend\'ariz-Pic\'on, V.~Mukhanov, and P.~J.~Steinhardt, 
Phys. \ Rev. \ Lett. {\bf 85}, 4438 (2000).
	
\bibitem{fR} 
S.~Capozziello, V.~F.~Cardone, S.~Carloni and A.~Troisi,
Int.\ J.\ Mod.\ Phys.\ D \textbf{12}, 1969 (2003); S.~M.~Carroll,
V.~Duvvuri, M.~Trodden and M.~S.~Turner, Phys.\ Rev.\ D 70,
043528 (2004);
S.~Nojiri and S.~D.~Odintsov,
Phys.\ Rev.\  D {\bf 68}, 123512 (2003).
	
\bibitem{st} 
J.~P.~Uzan,
Phys.\ Rev.\  D {\bf 59}, 123510 (1999); 
L.~Amendola,
Phys.\ Rev.\ D \textbf{60}, 043501 (1999); 
T.~Chiba, Phys.\ Rev.\ D \textbf{60}, 083508 (1999); 
N.~Bartolo and M.~Pietroni,
Phys.\ Rev.\  D {\bf 61}, 023518 (2000); 
F.~Perrotta, C.~Baccigalupi and
S.~Matarrese, Phys.\ Rev.\ D \textbf{61}, 023507 (2000); 
A.~Riazuelo and J.~P.~Uzan,
Phys.\ Rev.\ D {\bf 66}, 023525 (2002);
D.~F.~Torres,
Phys.\ Rev.\  D {\bf 66}, 043522 (2002);
E.~Elizalde, S.~Nojiri and S.~D.~Odintsov,
Phys.\ Rev.\  D {\bf 70}, 043539 (2004);
L.~Perivolaropoulos,
JCAP {\bf 0510}, 001 (2005);
D.~A.~Easson,
JCAP {\bf 0702}, 004 (2007);
S.~Nesseris and L.~Perivolaropoulos,
Phys.\ Rev.\  D {\bf 75}, 023517 (2007).

\bibitem{brane} 
G.~R.~Dvali, G.~Gabadadze and M.~Porrati,
Phys.\ Lett.\ B {\bf 485}, 208 (2000);
V.~Sahni and Y.~Shtanov,
JCAP {\bf 0311}, 014 (2003).

\bibitem{fRper}
P.~Zhang, 
Phys.\ Rev.\  D \textbf{73}, 123504 (2006);
S.~M.~Carroll, I.~Sawicki, A.~Silvestri and M.~Trodden, 
New J.\ Phys.\  \textbf{8}, 323 (2006); 
R.~Bean, D.~Bernat, L.~Pogosian, A.~Silvestri and M.~Trodden, 
Phys.\ Rev.\  D \textbf{75}, 064020 (2007);
Y.~S.~Song, W.~Hu and I.~Sawicki,
Phys.\ Rev.\  D {\bf 75}, 044004 (2007);
B.~Li and J.~D.~Barrow, 
Phys.\ Rev.\  D \textbf{75}, 084010 (2007);
T.~Faulkner, M.~Tegmark, E.~F.~Bunn and Y.~Mao,
arXiv:astro-ph/0612569;
I.~Sawicki and W.~Hu,
arXiv:astro-ph/0702278;
K.~Uddin, J.~E.~Lidsey and R.~Tavakol,
arXiv:0705.0232 [gr-qc];
S.~Tsujikawa,
arXiv:0709.1391 [astro-ph].

\bibitem{AKS}
L.~Amendola, M.~Kunz and D.~Sapone,
arXiv:0704.2421 [astro-ph].	

\bibitem{lensing}	
A.~R.~Cooray and D.~Huterer, 
Astrophys.\ J.\  {\bf 513}, 95 (1999);
N.~Sarbu, D.~Rusin and C.~P.~Ma, 
Astrophys.\ J.\  {\bf 561}, 147 (2001);
K.~H.~Chae {\it et al.}, 
Phys. Rev. Lett. {\bf 89}, 151301 (2002);
E.~V.~Linder, Phys. Rev. D {\bf 70},  043534 (2004);
W.~Hu and B.~Jain, Phys. Rev. D {\bf 70}, 
043009 (2004); 
D.~Jain, J.~S.~Alcaniz and A.~Dev,
Nucl.\ Phys.\  B {\bf 732}, 379 (2006);
M.~Takada and B.~Jain,
Mon.\ Not.\ Roy.\ Astron.\ Soc.\  {\bf 348}, 897 (2004);
J.~Albert {\it et al.}  [SNAP Collaboration],
arXiv:astro-ph/0507460;
A.~Refregier {\it et al.},
arXiv:astro-ph/0610062.

\bibitem{LSS}	
M.~Tegmark {\it et al.}  [SDSS Collaboration],
Phys.\ Rev.\  D {\bf 69}, 103501 (2004);
K.~Abazajian {\it et al.}  [SDSS Collaboration],
Astron.\ J.\  {\bf 128}, 502 (2004);
U.~Seljak {\it et al.}  [SDSS Collaboration], 
Phys.\ Rev.\ D {\bf 71}, 103515 (2005); 
M.~Tegmark {\it et al.},
Phys.\ Rev.\  D {\bf 74}, 123507 (2006).

\bibitem{Shirata}	
A.~Shirata, T.~Shiromizu, N.~Yoshida and Y.~Suto,
Phys.\ Rev.\  D {\bf 71}, 064030 (2005).

\bibitem{Lahav}	
O.~Lahav, P.~B.~Lilje, J.~R.~Primack and M.~J.~Rees,
Mon.\ Not.\ Roy.\ Astron.\ Soc.\  {\bf 251}, 128 (1991).

\bibitem{WS}	
L.~M.~Wang and P.~J.~Steinhardt,
Astrophys.\ J.\  {\bf 508}, 483 (1998).

\bibitem{Linder}
E.~V.~Linder,
Phys.\ Rev.\  D {\bf 72}, 043529 (2005);
D.~Huterer and E.~V.~Linder,
Phys.\ Rev.\  D {\bf 75}, 023519 (2007);
E.~V.~Linder and R.~N.~Cahn,
arXiv:astro-ph/0701317.

\bibitem{Ameper}	
L.~Amendola,
Phys.\ Rev.\ Lett.\  {\bf 93}, 181102 (2004).

\bibitem{Ame}
L.~Amendola,
Phys.\ Rev.\  D {\bf 62}, 043511 (2000).	

\bibitem{Esp}	
G.~Esposito-Farese and D.~Polarski,
Phys.\ Rev.\  D {\bf 63}, 063504 (2001).

\bibitem{GPRS}
R.~Gannouji, D.~Polarski, A.~Ranquet and A.~A.~Starobinsky,
JCAP {\bf 0609}, 016 (2006).

\bibitem{AGPT}
L.~Amendola, R.~Gannouji, D.~Polarski and S.~Tsujikawa,
Phys.\ Rev.\  D {\bf 75}, 083504 (2007).
	
\bibitem{Tsure}
S.~Tsujikawa,
Phys.\ Rev.\  D {\bf 72}, 083512 (2005).	
	
\bibitem{Hwang05}
J.~c.~Hwang and H.~Noh,
Phys.\ Rev.\  D {\bf 71}, 063536 (2005);
Phys.\ Rev.\  D {\bf 66}, 084009 (2002).	

\bibitem{Boi}	
B.~Boisseau, G.~Esposito-Farese, D.~Polarski 
and A.~A.~Starobinsky,
Phys.\ Rev.\ Lett.\  {\bf 85}, 2236 (2000).	

\bibitem{Schimd}
C.~Schimd, J.~P.~Uzan and A.~Riazuelo,
Phys.\ Rev.\  D {\bf 71}, 083512 (2005).

\bibitem{BD}
C.~Brans and R.~H.~Dicke,
Phys.\ Rev.\  {\bf 124}, 925 (1961).
	
\bibitem{Damour}
T.~Damour and K.~Nordtvedt,
Phys.\ Rev.\  D {\bf 48}, 3436 (1993).	

\bibitem{BDcon}
B.~Bertotti, L.~Iess and P.~Tortora,
Nature {\bf 425}, 374 (2003).

\bibitem{APT}
L.~Amendola, D.~Polarski and S.~Tsujikawa,
Phys.\ Rev.\ Lett.\  {\bf 98}, 131302 (2007).

\bibitem{Chiba}
T.~Chiba,
Phys.\ Lett.\  B {\bf 575}, 1 (2003).

\bibitem{AT}
L.~Amendola and S.~Tsujikawa,
arXiv:0705.0396 [astro-ph].

\bibitem{LGCrecent}
G.~J.~Olmo,
Phys.\ Rev.\ Lett.\  {\bf 95}, 261102 (2005);
J.~A.~R.~Cembranos,
Phys.\ Rev.\  D {\bf 73}, 064029 (2006);
A.~L.~Erickcek, T.~L.~Smith and M.~Kamionkowski,
Phys.\ Rev.\  D {\bf 74}, 121501 (2006);
V.~Faraoni,
Phys.\ Rev.\  D {\bf 74}, 023529 (2006);
A.~F.~Zakharov, A.~A.~Nucita, F.~De Paolis and G.~Ingrosso, 
Phys.\ Rev.\  D \textbf{74}, 107101 (2006); 
I.~Navarro and K.~Van Acoleyen,
JCAP {\bf 0702}, 022 (2007);
G.~Allemandi and M.~L.~Ruggiero,
arXiv:astro-ph/0610661;
X.~H.~Jin, D.~J.~Liu and X.~Z.~Li,
arXiv:astro-ph/0610854;
T.~Chiba, T.~L.~Smith and A.~L.~Erickcek,
arXiv:astro-ph/0611867;
P.~J.~Zhang,
arXiv:astro-ph/0701662;
W.~Hu and I.~Sawicki,
arXiv:0705.1158 [astro-ph].

\bibitem{tach}
A.~Sen,
JHEP {\bf 0204}, 048 (2002);
T.~Padmanabhan,
Phys.\ Rev.\  D {\bf 66}, 021301 (2002);
E.~J.~Copeland, M.~R.~Garousi, M.~Sami and S.~Tsujikawa,
Phys.\ Rev.\  D {\bf 71}, 043003 (2005).

\bibitem{Maeda}
K.~i.~Maeda,
Phys.\ Rev.\  D {\bf 39}, 3159 (1989).

\bibitem{Lucaper}
L.~Amendola,
Phys.\ Rev.\  D {\bf 69}, 103524 (2004).

\bibitem{Tsuji}
F.~Piazza and S.~Tsujikawa,
JCAP {\bf 0407}, 004 (2004);
S.~Tsujikawa and M.~Sami,
Phys.\ Lett.\  B {\bf 603}, 113 (2004).

\bibitem{AQTW}
L.~Amendola, M.~Quartin, S.~Tsujikawa and I.~Waga,
Phys.\ Rev.\  D {\bf 74}, 023525 (2006).

\bibitem{Gas}
M.~Gasperini and G.~Veneziano,
Phys.\ Rept.\  {\bf 373}, 1 (2003).

\bibitem{runaway}
M.~Gasperini, F.~Piazza and G.~Veneziano,
Phys.\ Rev.\  D {\bf 65}, 023508 (2002);
T.~Damour, F.~Piazza and G.~Veneziano,
Phys.\ Rev.\ Lett.\  {\bf 89}, 081601 (2002).

\bibitem{GB0}
S.~Nojiri, S.~D.~Odintsov and M.~Sasaki,
Phys.\ Rev.\  D {\bf 71}, 123509 (2005); 
M.~Sami, A.~Toporensky, P.~V.~Tretjakov and S.~Tsujikawa,
Phys.\ Lett.\ B {\bf 619}, 193 (2005); 
G.~Calcagni, S.~Tsujikawa and M.~Sami,
Class.\ Quant.\ Grav.\ {\bf 22}, 3977 (2005); 
S.~Nojiri, S.~D.~Odintsov and M.~Sami,
Phys.\ Rev.\  D {\bf 74}, 046004 (2006);
S.~Tsujikawa,
Annalen Phys.\  {\bf 15}, 302 (2006);
B.~M.~N.~Carter and I.~P.~Neupane, JCAP {\bf 0606}, 004 (2006);
T.~Koivisto and D.~F.~Mota, Phys.\ Lett.\  B {\bf 644}, 104 (2007);
Phys.\ Rev.\  D {\bf 75}, 023518 (2007);
S.~Tsujikawa and M.~Sami, JCAP {\bf 0701}, 006 (2007); 
B.~M.~Leith and I.~P.~Neupane, arXiv:hep-th/0702002.

\bibitem{GB1}
L.~Amendola, C.~Charmousis and S.~C.~Davis,
JCAP {\bf 0612}, 020 (2006).

\bibitem{GB2}
L.~Amendola, C.~Charmousis and S.~C.~Davis,
arXiv:0704.0175 [astro-ph].

\bibitem{instability}
A.~De Felice, M.~Hindmarsh and M.~Trodden,
JCAP {\bf 0608}, 005 (2006); G.~Calcagni, B.~de Carlos and A.~De Felice,
Nucl.\ Phys.\  B {\bf 752}, 404 (2006); Z.~K.~Guo, N.~Ohta and
S.~Tsujikawa, Phys.\ Rev.\  D {\bf 75}, 023520 (2007).

\end{thebibliography}
\end{document}